\documentclass[letter]{aa}

\usepackage{graphicx}
\usepackage{subfigure}
\usepackage{caption} 
\usepackage{tabularx}
\usepackage{txfonts}
\usepackage{fontawesome}
\usepackage{hyperref}
\usepackage{xcolor}
\usepackage[acronym]{glossaries}

\newacronym{PSF}{PSF}{Point Spread Function}
\newacronym{MSE}{MSE}{Mean Square Error}
\newacronym{SRA}{SRA}{Sparse Reconstruction Algorithm}
\newacronym{SCORE}{SCORE}{Shape COnstraint REstoration algorithm}
\newacronym{PSNR}{PSNR}{Peak Signal to Noise Ratio}
\newacronym{SNR}{SNR}{Signal to Noise Ratio}
\newacronym{T-RECS}{T-RECS}{Tiered Radio Extragalactic Continuum Simulation}
\newacronym{MEM}{MEM}{Maximum Entropy Method}
\newacronym{DL}{DL}{Deep Learning}
\newacronym{SKA}{SKA}{Square Kilometre Array}
\newacronym{DIDN}{DIDN}{Deep Iterative Down-Up CNN for Image Denoising}
\newacronym{HST}{HST}{Hubble Space Telescope}
\newacronym{CFHT}{CFHT}{Canada France Hawaii Telescope}

\begin{document}

   \title{ShapeNet: Shape Constraint for Galaxy Image Deconvolution}
   

   \author{F. Nammour
          \inst{1}
          \and
          U. Akhaury
          \inst{2}
          \and
          J. N. Girard
          \inst{3}
          \and
          F. Lanusse
          \inst{1}
          \and
          F. Sureau
          \inst{4}
          \and
          C. Ben Ali
          \inst{5}
          \and
          J.-L. Starck
          \inst{1}
          }

   \institute{AIM, CEA, CNRS, Universit\'e Paris-Saclay, Universit\'e de Paris, F-91191 Gif-sur-Yvette, France.\\
              \email{fadinammour95@gmail.com}\\
              \email{francois.lanusse@cea.fr}\\
              \email{jstarck@cea.fr}
         \and
             Laboratoire d’astrophysique, École Polytechnique Fédérale de Lausanne (EPFL), Switzerland.\\
             \email{utsav.akhaury@epfl.ch}
        \and
             LESIA, Observatoire de Paris, Université PSL, CNRS, Sorbonne Université, Université de Paris, 5 place Jules Janssen, 92195 Meudon, France.\\
             \email{julien.girard@obspm.fr}
        \and
             Universit\'e Paris-Saclay, CEA, CNRS, Inserm, BioMAPs, 91401 Orsay, France.\\
             \email{florent.sureau@cea.fr}
        \and
             Wisear, 11 rue des Cassoirs, 89000 Auxerre, France.\\
             \email{claire.benali3@gmail.com}
             }

    \date{}
   
 
  \abstract
   {\acrfull{DL} has shown remarkable results in solving inverse problems in various domains. In particular, the Tikhonet approach is very powerful to deconvolve optical astronomical images \citep{sureau2020}. Yet, this approach only uses the $\ell_2$ loss, which does not guarantee the preservation of physical information (e.g. flux and shape) of the object reconstructed in the image. In \cite{nammour2021galaxy}, a new loss function was proposed in the framework of sparse deconvolution, which better preserves the shape of galaxies and reduces the pixel error. In this paper, we extend Tikhonet to take into account this shape constraint, and apply our new \acrshort{DL} method, called ShapeNet, to optical and radio-interferometry simulated data set. The originality of the paper relies on i) the shape constraint we use in the neural network framework, ii) the application of deep learning to radio-interferometry image deconvolution for the first time, and iii) the generation of a simulated radio data set that we make available for the community. A range of examples illustrates the results.}


   \maketitle
   
\section{Introduction}
\label{sec:intro}
Sparse wavelet regularization techniques, based either on the $\ell_0$ or $\ell_1$ norm, have been state-of-the art for astronomical image deconvolution for years, leading to striking results such as the improvement of the resolution by a factor 4 (2 in each dimension) on the Cygnus-A radio image reconstruction compared to the CLEAN standard algorithm \citep{garsden2015}. Sparsity, similarly to the positivity regularization constraint, can be considered as a weak prior on the distribution of the wavelet coefficients of the solution, as most if not all images present a compressible behaviour in the wavelet domain. In recent years, the emergence of \acrfull{DL} has shown promising results in various domains, including deconvolution \citep{xu2014dcnn}. In astrophysics, methods based on \acrshort{DL} have already been developed to perform model fitting which can be seen as a parametric deconvolution \citep{tuccillo2018deepfitting}. \citet{sureau2020} introduced the Tikhonet neural network for optical galaxy image deconvolution. 
Tikhonet clearly outperformed sparse regularization for both the \acrshort{MSE} and a shape criterion, the galaxy shape being encoded through a measure of its ellipticity \citep{sureau2020}. These great results can be explained by the fact that \acrshort{DL} learns, or rather approximates, the mean of the posterior distribution of the solution. As there is no guarantee that a non-linear deconvolution process preserves the galaxies’ shapes, \citet{nammour2021galaxy} introduced a new shape penalization term and showed that adding this penalty to sparse regularization improves both the solution shape and the MSE.
In this paper, we propose a new deconvolution method, called ShapeNet, by extending the Tikhonet method to include the shape constraints. We present first results for optical galaxies image deconvolution, then we show that both Tikhonet and ShapeNet can also be used in the framework of radio galaxy image deconvolution. To achieve these results, 
we have developed our own datasets for both optical and radio astronomical images. The optical dataset generation uses \acrshort{HST}-like target images and simulates \acrshort{CFHT}-like noisy observations by using real images and PSFs from the COSMOS catalog \citep{cosmos2012mandelbaum}. The radio dataset comprises noisy images with realistic PSFs similar to those of the MeerKAT telescope, and parametric galaxies with properties taken from the \acrshort{T-RECS} catalog \citep{bonaldi2019}.The optical and radio dataset generation are adapted for Machine Learning, and are explained in \ref{subsec:optical_data} and \ref{subsec:radio_data} respectively. 
Section~2 introduces our new methodology and Section~3 presents the results of numerical experiments. We conclude in section~4.


\section{Deep learning deconvolution with a shape constraint}
\label{sec:methodology}

\subsection{The deconvolution problem}
 Denoting the observed image by $y\in\mathbb{R}^{n\times n}$ and the PSF by $h\in\mathbb{R}^{n\times n}$, the observational model is as follows:
\begin{equation}
 \label{eq:invprob}
  y = h \ast x_T + n \quad,
\end{equation}
where $x_T \in \mathbb{R}^{n \times n}$ is the ground truth image, and $n \in\mathbb{R}^{n\times n}$ is additive noise. We can partially restore $y$ by applying the least squares method. In this case, the solution oscillates because the problem in eq.~\ref{eq:invprob} is ill-conditioned. More generally, it is an ill-posed problem and can instead, be tackled using regularization \citep{bertero1998introduction}. For example, by noting $H\in\mathbb{R}^{n^2 \times n^2}$ the circulant matrix corresponding to the convolution operator $h$, the Tikhonov solution of eq.~\ref{eq:invprob} is:
\begin{equation}
    \label{eq:tikhonov}
    \hat{x} = \left(H^\top H+\lambda \Gamma^\top \Gamma\right)^{-1}H^\top y\quad,
\end{equation}
where $\Gamma\in\mathbb{R}^{n^2 \times n^2}$ is the Tikhonov linear filter, and $\lambda\in\mathbb{R}_+$ is the regularization weight. 

\subsection{Tikhonet Deconvolution}
\label{subsec:tikho}
The Tikhonet is a two steps \acrshort{DL} approach to solve deconvolution problems. 
The first step is to perform deconvolution using a Tikhonov filter with a quadratic regularisation
and setting $\Gamma = {\mathrm{Id}}$, leading to a deconvolved image containing correlated additive noise which is filtered in a second step 
 using a 4 scales \emph{XDense U-Net} \citep{sureau2020}. The network is trained to learn the mapping between the Tikhonov output and the target image using the \acrfull{MSE} as a loss function.
The regularisation weight is estimated for each image using a SURE risk minimisation with an estimate of the image \acrshort{SNR}.

\subsection{The Shape Constraint}
\label{subsec:shape_constraint}
The shape information of galaxies is essential in various fields of astrophysics, such as Galaxy Evolution and Cosmology. The measure that is used to study the shape of a galaxy is the ellipticity, which is a complex scalar, $e = e_1 + \bold{i} e_2$. The ellipticity of an image $x$ is given by \citep{kaiser1995ksb}:
\begin{equation}
    e(x) = \frac{\mu_{2,0}(x)-\mu_{0,2}(x) + \bold{i} 2\mu_{1,1}(x)}{\mu_{2,0}(x)+\mu_{0,2}(x)} \quad,
\end{equation}
where $\mu_{s,t}$ are the image centered moments of order $(s+t)$ defined as:
\begin{equation}
    \mu_{s,t}(x) = \sum_{i=1}^n \sum_{j=1}^n x\left[(i-1)n+j\right]\left(i-i_c\right)^s\left(j-j_c\right)^t \quad,
\end{equation}
and $i_c$ and $j_c$ are the coordinated of the centroid of $x$, such that:
\begin{equation}
    i_c = \frac{\sum_{i=1}^n \sum_{j=1}^n i\cdot x\left[(i-1)n+j\right]}{\sum_{i=1}^n \sum_{j=1}^n x\left[(i-1)n+j\right]}
\end{equation}
and
\begin{equation}
    j_c = \frac{\sum_{i=1}^n \sum_{j=1}^n j\cdot x\left[(i-1)n+j\right]}{\sum_{i=1}^n \sum_{j=1}^n x\left[(i-1)n+j\right]} \quad.
\end{equation}


In \cite{nammour2021galaxy}, we derived the following reformulation:
\begin{equation}
    \label{eq:e1_scal}
    e_1 = \frac{\left<x,u_3\right>\left<x,u_4\right>-\left<x,u_1\right>^2+\left<x,u_2\right>^2}{\left<x,u_3\right>\left<x,u_4\right>-\left<x,u_1\right>^2-\left<x,u_2\right>^2}
\end{equation}
and
\begin{equation}
    \label{eq:e2_scal}
    e_2 = \frac{2\left(\left<x,u_3\right>\left<x,u_6\right>-\left<x,u_1\right>\left<x,u_2\right>\right)}{\left<x,u_3\right>\left<x,u_4\right>-\left<x,u_1\right>^2-\left<x,u_2\right>^2} \quad,
\end{equation}
where $\left\{u_1,\dots,u_6\right\} \in\mathbb{R}^{6 \times n \times n}$ are constant images defined, for all $i$, $j$ in $\{1,\cdots, n\}$, as:
\begin{equation}
    \begin{aligned}
        &u_1[(i-1)n+j] = i, &&u_2[(i-1)n+j] = j,\\
        &u_3[(i-1)n+j] = 1, &&u_4[(i-1)n+j] = (i^2+j^2),\\
        &u_5[(i-1)n+j] = (i^2-j^2), &&u_6[(i-1)n+j] = (ij).
    \end{aligned}
\end{equation}
All the scalar products in eq. \ref{eq:e1_scal} and \ref{eq:e2_scal} are linear in $x$. Therefore, by formulating the shape constraint as a data-fidelity term in these scalar products space, we obtain:
\begin{equation}
    \label{eq:M0}
    M_0(x) = \sum_{i=1}^6 \omega_i \left<h \ast x - y, u_i\right>^2 \quad,
\end{equation}
where $\left\{\omega_1,\dots,\omega_6\right\}$ are non-negative scalar weights. Eq. \ref{eq:M0} offers a shape constraint which properties are straightforwardly derived. However the ellipticity measure is extremely sensitive regarding noise. To add robustness to the shape constraint, we considered windowing the observed image, $y$, to reduce the noise effect. One approach is to fit a Gaussian window on $y$, however it would require an additional preprocessing step on each observed image. To avoid this step, we chose to use a set of precomputed windows such that at least one of them fits the galaxy in the observed image. We considered curvelets to look for a candidate set, since they are a family of linear multi-scale transforms that are usually designed with specific properties to efficiently represent objects of interest. This lead us to:
\begin{equation}
    \label{eq:shape_constraint}
    M(x) = \sum_{i=1}^6 \sum_{j=1}^K \omega_{ij} \left<\psi_j\left(h\ast x -y\right),u_i\right>^2\quad,
\end{equation}
where $\left\{\omega_{ij}\right\}_{i,j}$ are non-negative scalar weights (their computation is detailed in \cite{nammour2021galaxy}) and 
$\left\{\psi_j\right\}_{j}$ are  $K$ directional and multi-scale filters, derived from a 
curvelet-like decomposition \citep{starck2015sparse,kutyniok2012shearlets}.  These filters allow to capture the anisotropy of the galaxy image and are used in the constraint as a set of windows such that at least one them reduces the noise in the image and emphasize the useful signal. We have also shown that adding such a constraint to a sparse deconvolution approach reduces both the shape and pixel errors \citep{nammour2021galaxy}.
An alternative could have been to directly use the ellipticity in the loss function rather than our shape constraint. However, it would have raised some serious issues. The ellipticity measurement is very sensitive to noise, and it would have been necessary to take into account the noise propagation on the ellipticity measurements. As the propagated noise would clearly not be Gaussian, this is far from being trivial. Furthermore, the ellipticity operator is non-linear in $x$, which complicates gradient computation, thus rendering optimisation to be much more difficult. On the contrary, our shape constraint is a sum of weighted-squared fully linear components and the noise can be well controlled.

\subsection{ShapeNet Deconvolution}
\label{subsec:shapenet}
This shape constraint could also be used in a deep learning deconvolution framework, by extending the Tikhonet method. We propose here the ShapeNet \acrshort{DL} deconvolution, applying the following updates to Tikhonet: 
\begin{itemize}
    \item We set $\Gamma$ to a Laplacian filter instead of the identity.  
     This is motivated by the fact that images have generally a decreasing profile in the Fourier space and the data quality at high frequencies is more deteriorated than at low ones. 
     \item The weight of the regularisation parameter $\lambda$ is constant for all images, which adds homogeneity to the filter in the ShapeNet pipeline and improves the explainability of the task learned by the XDense U-Net. 
     \item The loss function used for the training of the network contains an additional term, which is the shape constraint.
\end{itemize}
The loss function is now expressed as:
\begin{equation}
  l(\tilde{x}) = \|\tilde{x} - x_T\|_2^2 + \gamma M(\tilde{x}) \quad,
\end{equation}
where $\tilde{x}$ is the Tikhonov filter output of the observed image, $M$ is the shape constraint and $\gamma \in \mathbb{R_+}$ its weight parameter as introduced in \cite{nammour2021galaxy}. The choice of the hyper-parameters is detailed in \ref{subsubsec:implementation_optical} and \ref{subsubsec:implementation_radio}.
  
\section{Numerical Experiments \& Results}
\label{sec:experiments}
In this section, we present the numerical experiments carried out in order to assess the methods discussed above. The code was developed using \texttt{Python 3.7.5} and \texttt{TensorFlow 1.15.2}, \texttt{Keras 2.3.1} \citep{chollet2015keras}, \texttt{AlphaTransform}\footnote{\url{https://github.com/dedale-fet/alpha-transform/}}, Matplotlib 3.1.3 \citep{hunter2007matplotlib}, Galaxy2Galaxy and GalFlow \citep{lanusse2015galflow}. While training the U-Net for the \acrshort{DL} methods with and without shape constraint, we normalized the pixel values of the input images by $4\times 10^3$ for the optical case and by $2\times 10^3$ for the radio case in order to make their magnitudes close to unity, so that the activation functions in the neural network could better discriminate data. We compare the deep learning methods to \acrfull{SRA} and \acrfull{SCORE} \citep{nammour2021galaxy}, and additionally CLEAN for the radio case. \acrshort{SRA} is a deconvolution method based on sparsity and positivity, while \acrshort{SCORE} is its extension that uses an additional shape constraint, as described in \cite{nammour2021galaxy}. The qualitative criteria used for these experiments are:
\begin{itemize}
    \item \acrshort{MSE} $ =  \frac{\sum_k \left( x[k] - x_T[k] \right)^2}{\sum_k x_T[k]^2}$
    \item Relative error of the flux density: $\Delta F = \left|\frac{\sum_k  (x[k] - x_T[k])}{\sum_k x_T[k]}\right|$
    \item Mean absolute error of each component of the ellipticity: $e_1$ and $e_2$
\end{itemize}
The ellipticity in this case was estimated using the adaptive moments method \citep{hirata2003} which consists of fitting an elliptical Gaussian window on the input image and then deducing the measurement from the window obtained.

\subsection{Optical Experiments}
\label{subsec:optical_exp}
\begin{figure}
 			\centering
 			\includegraphics[width=\linewidth]{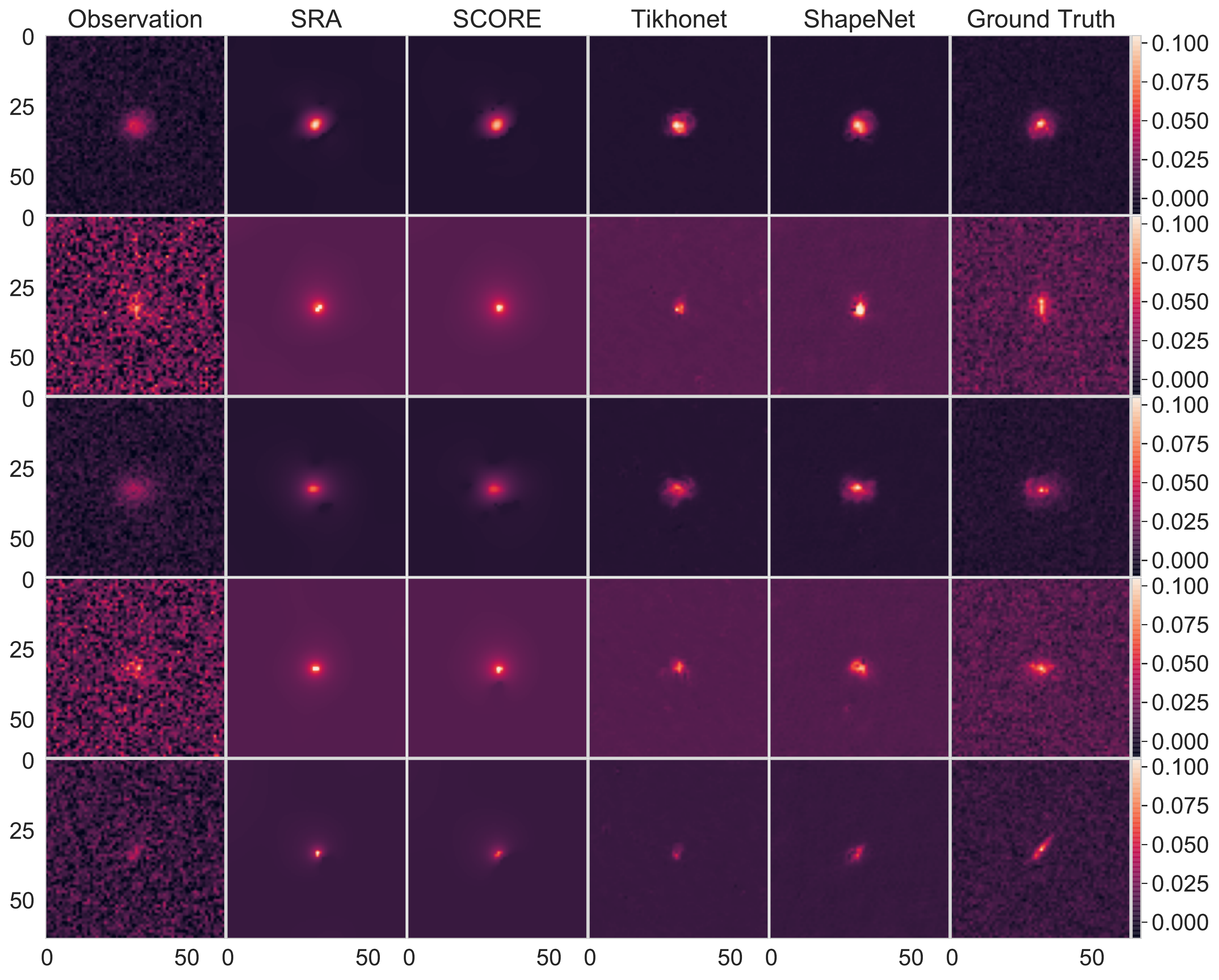}
 			\caption{Examples of extended galaxies reconstructed using the CFHT2HST dataset.}
 			\label{fig:cfht2hst_examples}
 		\end{figure}

    For the optical experiments, we used the CFHT2HST dataset, whose implementation is detailed in \ref{subsubsec:implementation_optical}. For visual comparison, we show 5 examples of resolved galaxies in figure \ref{fig:cfht2hst_examples}. Firstly, we observe that the sparse methods tend to make the reconstructed object more isotropic and smoother, noting that the fine details present in the ground truth are lost during reconstruction. This effect is explained by the use of starlets which are isotropic wavelets. The deep learning methods preserve more details and structures, as seen for the first galaxy. On adding the shape constraint, whether going from \acrshort{SRA} to \acrshort{SCORE} or from Tikhonet to ShapeNet, there is a better coincidence between the orientation of the reconstructed galaxies and the ground truth. We also notice that the addition of the shape constraint allows a better reconstruction of large galaxies. This can be seen for the Tikhonet results, where the galaxies reconstructed without the constraint appear to be smaller and less bright than the ground truth at a significant noise level. 

 	\begin{table}
 	    \resizebox{0.5\textwidth}{!}{
 		\begin{tabular}{|l||l|l|}
 			\hline
 			Methods & \acrshort{MSE}  & Flux\\
 			\hline
 			SRA & $4.33 \times 10^{-1} \pm 6.95 \times 10^{-3}$  & $2.06 \times 10^{0} \pm 5.92 \times 10^{-1}$\\
 			SCORE & $2.93 \times 10^{-1} \pm 3.93 \times 10^{-3}$  & $1.83 \times 10^{0} \pm 5.68 \times 10^{-1}$\\
 			Tikhonet & $3.33 \times 10^{-1} \pm 4.45 \times 10^{-3}$  & $1.25 \times 10^{0} \pm 4.16 \times 10^{-1}$\\
 			ShapeNet & $2.45 \times 10^{-1} \pm 3.47 \times 10^{-3}$  & $6.49 \times 10^{-1} \pm 6.20 \times 10^{-2}$\\
 			\hline
 			Methods & $e_1$ & $e_2$ \\ \hline
 			SRA & $1.28 \times 10^{-1} \pm 2.03 \times 10^{-3}$ & $1.54 \times 10^{-1} \pm 2.41 \times 10^{-3}$ \\
 			SCORE & $1.18 \times 10^{-1} \pm 1.84 \times 10^{-3}$ & $1.45 \times 10^{-1} \pm 2.20 \times 10^{-3}$ \\
 			Tikhonet & $1.24 \times 10^{-1} \pm 1.89 \times 10^{-3}$ & $1.34 \times 10^{-1} \pm 2.09 \times 10^{-3}$ \\
 			ShapeNet & $1.16 \times 10^{-1} \pm 1.81 \times 10^{-3}$ & $1.27 \times 10^{-1} \pm 2.06 \times 10^{-3}$ \\
 			\hline 
     	\end{tabular} \\
     	}
	 	\caption[CFHT2HST dataset results.]{CFHT2HST dataset results. The values indicate the mean error of each quantity and the uncertainty corresponds to the standard error of the given value.}
	 	\label{tab:cfht_results}
	 	\centering
 	\end{table}
 	
\begin{figure}
 		\centering
	 	\subfigure[Mean relative error of flux. ]
     	{\label{fig:cfht2hst_flux}
     	\includegraphics[width=\linewidth]
     	{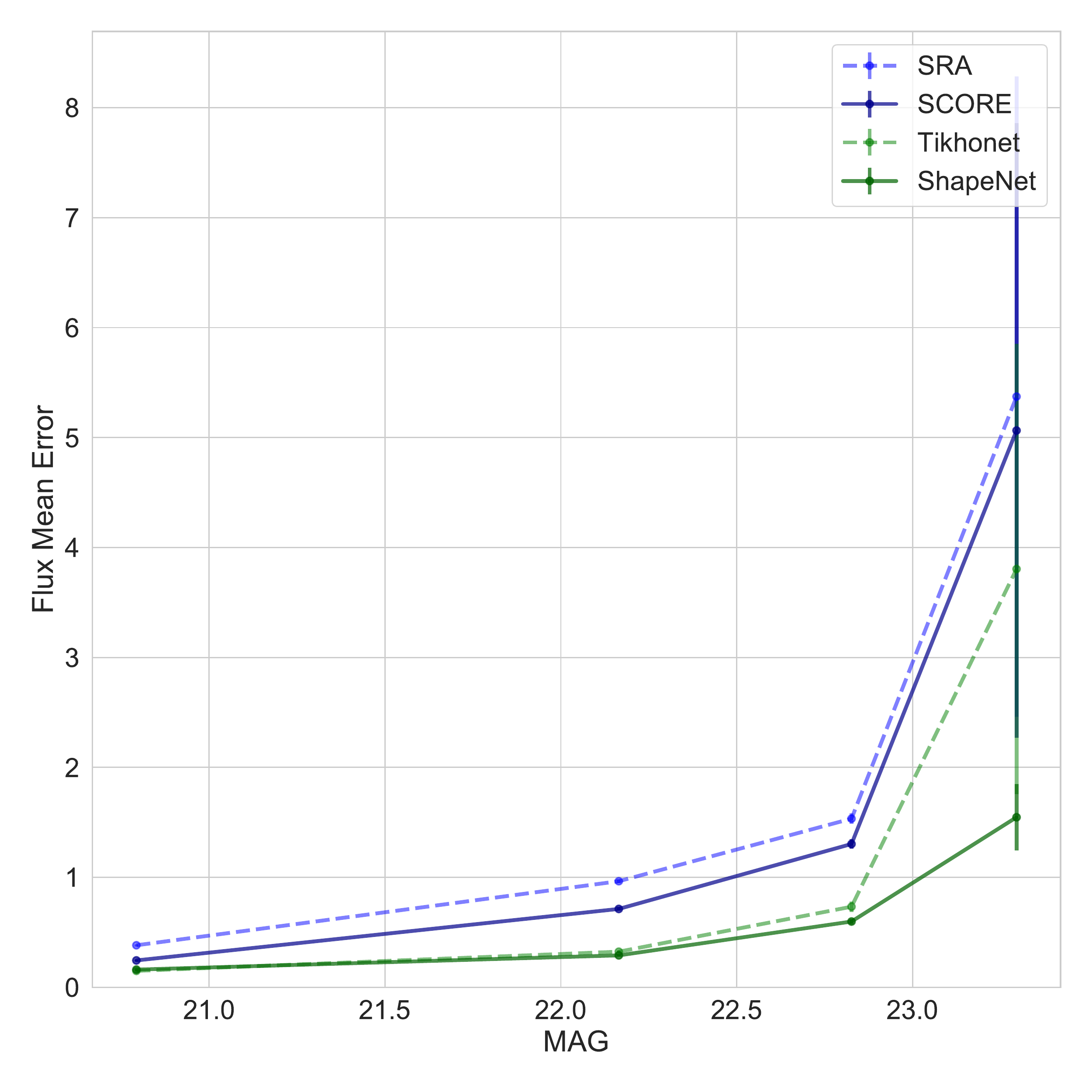}}%
     	\\
     	\subfigure[Mean error of $\|e\|_2$ (with $e=e_1+ie_2$).]
     	{\label{fig:cfht2hst_e}
     	\includegraphics[width=\linewidth]
     	{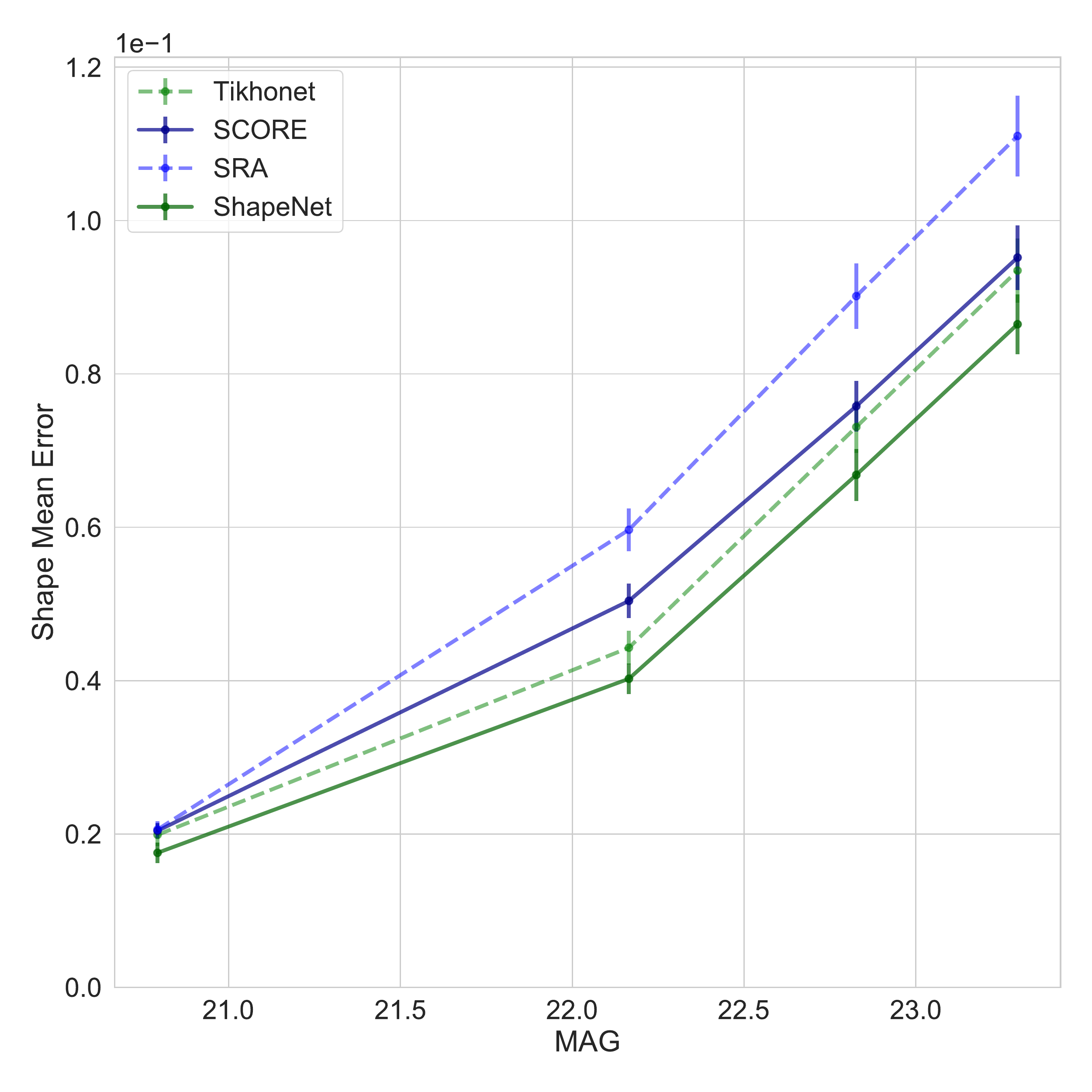}}%
     	\caption{Mean errors of reconstruction as a function of Magnitude for the CFHT2HST dataset, where the error bars correspond to the standard error of the given value.}
     	\label{fig:cfht2hst_plots}
 	\end{figure}

    To measure the \acrshort{MSE}, we carried out a weighting with the elliptical Gaussian window obtained while estimating the ground truth galaxy shape. This weighting reduces the noise in the ground truth images, allowing to reduce the bias in the estimation of \acrshort{MSE}. Quantitatively, we also compared sparse methods to deep learning methods by taking \acrshort{SRA} and Tikhonet. Our results corroborate those   of \cite{sureau2020}, i.e. in almost all measurements carried out in Table \ref{tab:cfht_results} and in figure \ref{fig:cfht2hst_plots}, Tikhonet average errors are lower than those obtained with \acrshort{SRA}.
   We notice that for both methodologies, sparsity and neural network, adding the shape constraint reduces the errors.
   From table \ref{tab:cfht_results} again, the ShapeNet reconstructions have lower errors than those obtained with Tikhonet, with reductions of 26\%, 48\%, 6\%, and 5\% for \acrshort{MSE}, flux, $ e_1 $ and $ e_2 $ respectively. Thus, we show that the addition of the shape constraint improves the performance of the methods considered with respect to all the quality criteria studied, which corroborates the results in \citet{nammour2021galaxy}.

    In conclusion, the experiments carried out on the CFHT2HST dataset show that adding the shape constraint in a deep learning framework significantly reduces the reconstruction errors.
    
\subsection{Radio Experiments}
\label{subsec:radio_exp}
\begin{figure}
 			\centering
 			\includegraphics[width=\linewidth]{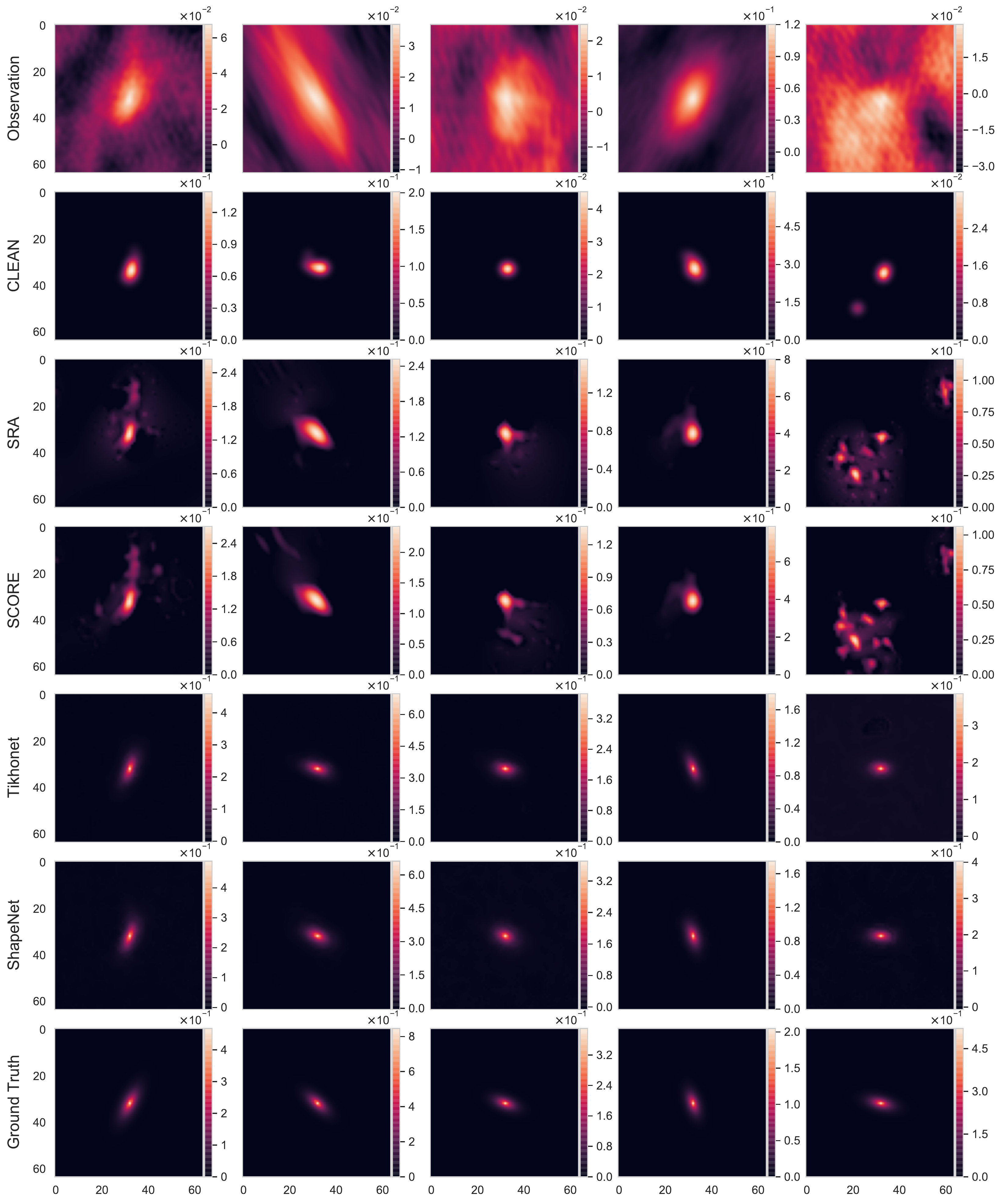}
 			\caption{Examples of galaxies, with \acrshort{PSNR} greater than 3, reconstructed from the \mbox{MeerKAT3600} dataset.}
 			\label{fig:meerkat3600_examples}
 		\end{figure}

    So far, deep learning deconvolution methods have only been tested on optical data. In this section, we extend the results of \citet{sureau2020}, by first investigating how deep learning techniques perform for radio galaxy deconvolution, and then adding a shape constraint during the neural network training. For the radio experiments, we used the MeerKAT3600 dataset, whose implementation is detailed in \ref{subsubsec:implementation_radio}. In figure \ref{fig:meerkat3600_examples}, we show examples of galaxies where the \acrshort{PSNR} is greater than 3, such that we can distinguish galaxies from noise in observations. Both CLEAN and sparse recovered images
    are smoother than the corresponding ground truths, and the orientation of the reconstructed galaxies is strongly biased by that of the \acrshort{PSF}. Deep learning methods better preserve the details such as shape, size, and orientation. Notice that the deep learning methods are able to detect galaxies even when the \acrshort{PSNR} is extremely low (as seen in the last column of figure \ref{fig:meerkat3600_examples}). Furthermore, the shape constraint helps to noticeably improve the results for galaxies with \acrshort{PSNR} greater than 10. 
    
    Similar to the optical case, adding the shape constraint to both sparse and deep learning methods improves their performance in the radio case as well. In more detail, from table \ref{tab:meerkat_results}, the ShapeNet reconstructions have lower errors than those obtained with Tikhonet, with reductions of 17\%, 7\%, 1\%, and 2\% for \acrshort{MSE}, flux, $ e_1 $ and $ e_2 $ respectively. 
    
    
    Eventually, we conclude that the sparsity and deep learning methods have better performance than CLEAN, adding the shape constraint brings a gain in all the quality criteria considered, and deep learning offers better performance than sparsity. Finally, ShapeNet outperforms all other methods discussed above.
    
     	\begin{table}
     	    \resizebox{0.5\textwidth}{!}{
     		\begin{tabular}{|l||l|l|}
     			\hline
     			Methods & MSE & Flux\\
     			\hline
     			CLEAN & $3.91 \times  10^{-1} \pm 3.51 \times  10^{-3}$  & $5.02 \times  10^{-1} \pm 3.47 \times  10^{-3}$\\
     			SRA & $2.81 \times  10^{-1} \pm 1.16 \times  10^{-2}$  & $3.31 \times  10^{-1} \pm 9.30 \times  10^{-3}$\\
     			SCORE & $2.69 \times  10^{-1} \pm 1.08 \times  10^{-2}$  & $2.39 \times  10^{-1} \pm 7.15 \times  10^{-3}$\\
     			Tikhonet & $5.21 \times  10^{-2} \pm 1.47 \times  10^{-3}$  & $1.58 \times  10^{-1} \pm 2.14 \times  10^{-3}$\\
     			ShapeNet & $4.33 \times  10^{-2} \pm 1.17 \times  10^{-3}$  & $1.47 \times  10^{-1} \pm 2.48 \times  10^{-3}$\\
     			\hline
     			Methods & $e_1$ & $e_2$ \\
     			\hline
     			CLEAN & $1.30 \times 10^{-1} \pm 1.76 \times  10^{-3}$ & $1.27 \times  10^{-1} \pm 1.70 \times  10^{-3}$ \\
     			SRA & $1.20 \times  10^{-1} \pm 1.87 \times  10^{-3}$ & $1.26 \times  10^{-1} \pm 2.07 \times  10^{-3}$ \\
     			SCORE & $1.14 \times  10^{-1} \pm 1.78 \times  10^{-3}$ & $1.21 \times  10^{-1} \pm 2.01 \times  10^{-3}$ \\
     			Tikhonet & $7.44 \times  10^{-2} \pm 1.44 \times  10^{-3}$ & $7.88 \times  10^{-2} \pm 1.53 \times  10^{-3}$ \\
     			ShapeNet & $7.34 \times  10^{-2} \pm 1.41 \times  10^{-3}$ & $7.70 \times  10^{-2} \pm 1.48 \times  10^{-3}$ \\
     			\hline
     		\end{tabular} \\
     		}
     		\caption[MeerKAT3600 dataset results.]{MeerKAT3600 dataset results. The values indicate the mean error of each quantity and the uncertainty corresponds to the standard error of the given value.}
     		\centering
     		\label{tab:meerkat_results}
     	\end{table}
     	
     	\begin{figure}
     		\centering
     		\label{fig:meerkat3600_plots}
     		\subfigure[Mean relative error of flux.]
     		{\label{fig:meerkat3600_flux}
     		\includegraphics[width=0.5\textwidth]
     		{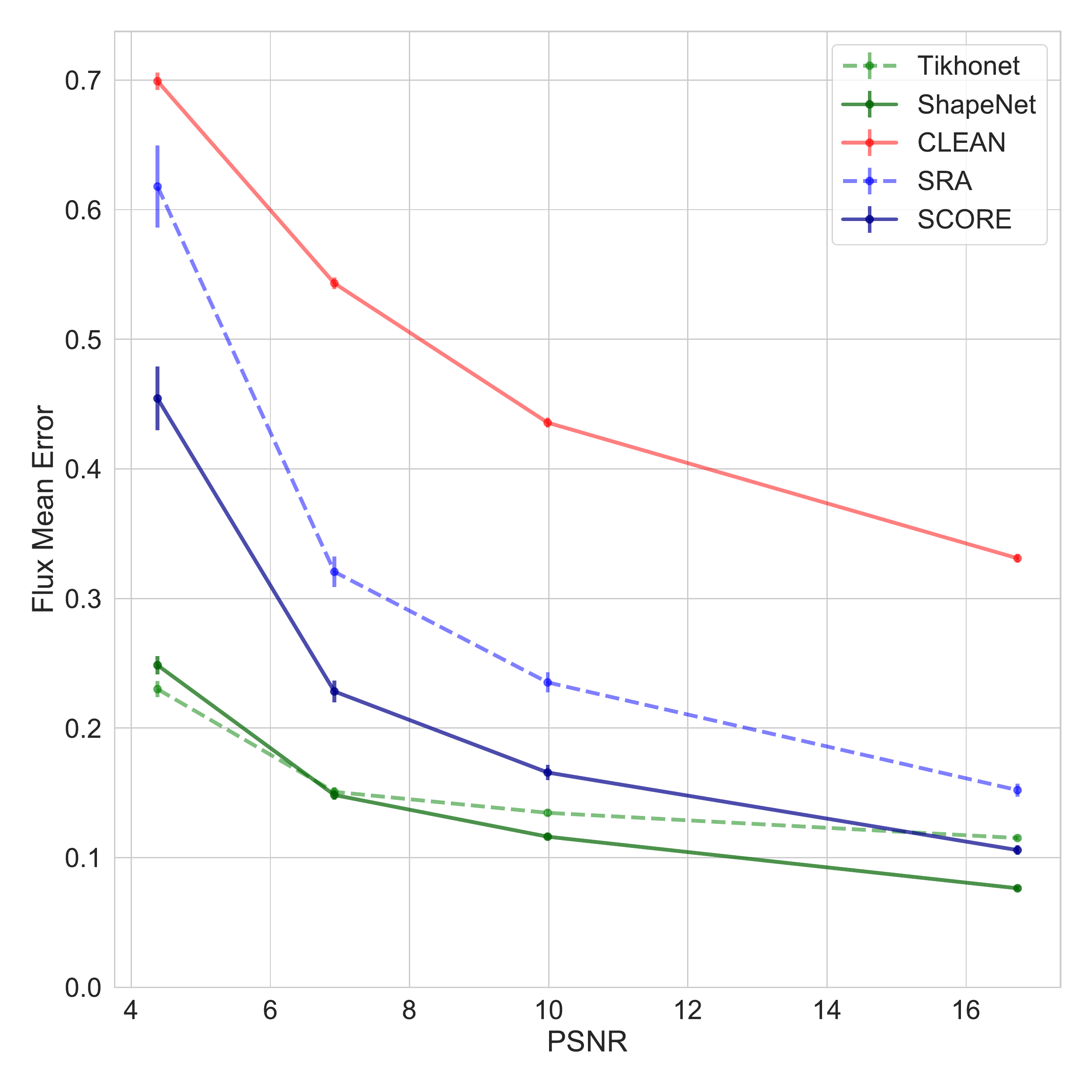}}%
     		\\
     		\subfigure[Mean error of $\|e\|_2$ (with $e=e_1+ie_2$).]
     		{\label{fig:meerkat3600_e}
     		\includegraphics[width=0.5\textwidth]
     		{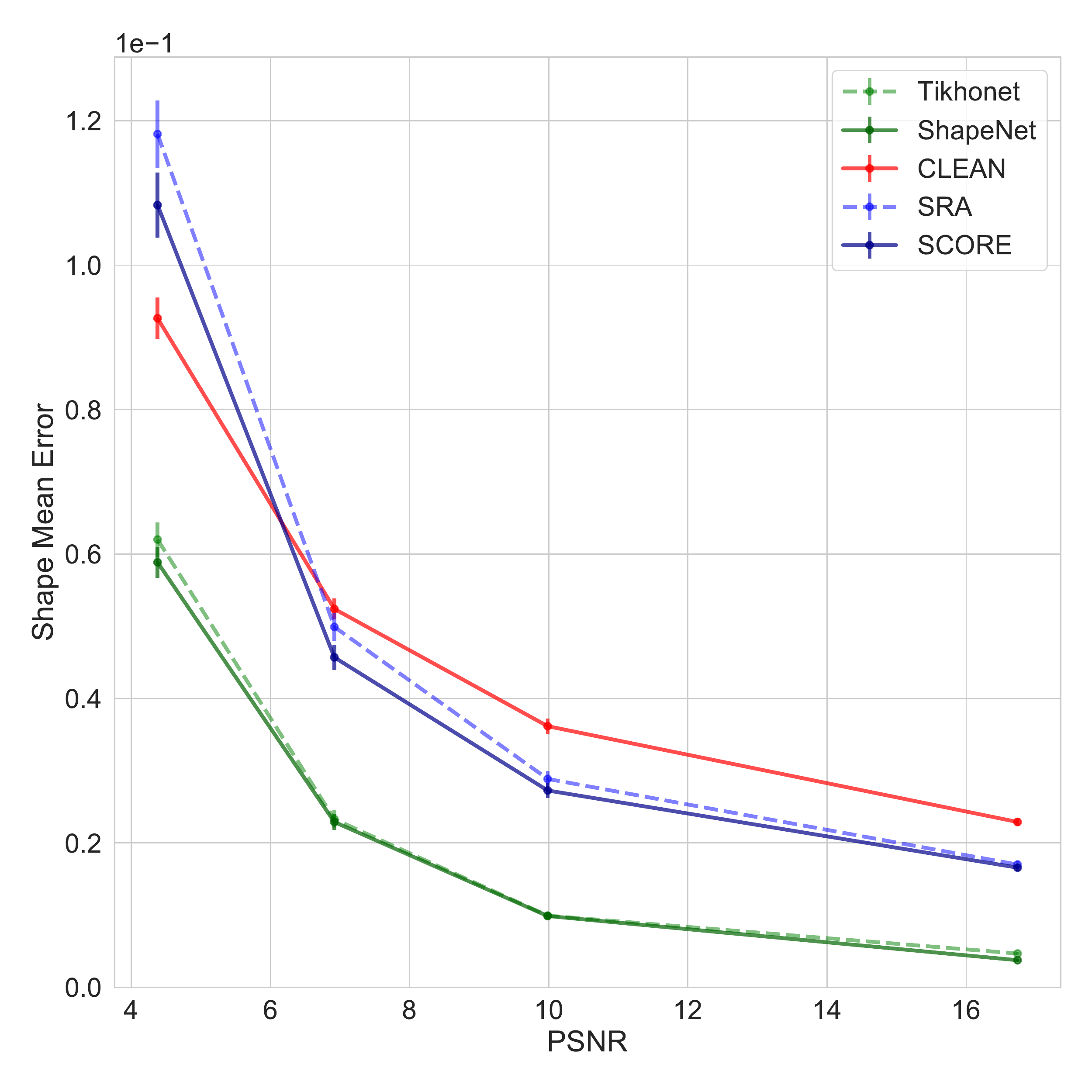}}%
     		\caption{Mean errors of reconstruction as a function of PSNR for the MeerKAT3600 dataset.}
            \label{fig:meerkat3600_plots}
     	 \end{figure}
     		
    
     		
     		

\section{Conclusion}
\label{sec:conclusion}
In this paper, we have introduced ShapeNet, a new problem specific approach, based on optimisation and \acrshort{DL}, to solve the galaxies deconvolution problem. 
We have developed and generated two realistic datasets, adapted for our numerical experiments, in particular the training step in \acrshort{DL}.
Our work extends the results of \citet{sureau2020}, by first investigating how deep learning techniques behave for radio-interferometry galaxy image deconvolution, and then adding a shape constraint during the neural network training.
Our experiments have shown that  both Tikhonet and 
 ShapeNet \acrshort{DL} deconvolution methods  allow us to better reconstruct radio-interferometry image reconstructions. 
We have shown that the shape constraint improves the performance of galaxy deconvolution, both for optical and radio-interferometry images, for different criteria such as the pixel error, the flux and the ellipticity. In practice, this method can be used with a source extraction algorithm to restore wide field images containing multiple galaxies.

The evaluation of our method on real data will be done in the future.
ShapeNet could be improved by replacing the U-Net by a more competitive denoiser such as \acrfull{DIDN} \citep{yu2019deep} or \citet{laine2019high} method. Additionally, the ShapeNet could also be improved by adding filters to extract feature maps before the deconvolution step in a similar fashion to \citet{dong2021deep}. We are currently investigating the addition of the shape constraint to an ameliorated version of the ADMMnet method presented in \citet{sureau2020}, and the amelioration concerns the properties of the neural network which are discussed in \citet{pesquet2021mmo}. On a wider scope, our deconvolution method can also be applied to other fields that will be addressed by \acrshort{SKA}. An efficient deconvolution method, accessible to the community, enables reconstructing a better estimate of the sky with fewer data than classical methods. This is key for optimizing both the observing time and the number of required data to achieve a given image reconstruction fidelity. In the upcoming surveys involving SKA1-MID, the use of deep-learning methods that are specific to the behaviour of an instrument during an observation will be critical to limit the deluge of data produced by the new generation of radio interferometers. 



\begin{acknowledgements}
      We would like to thank Axel Guinot and Hippolyte Karakostanoglou for their valuable help implementing the optical dataset.
\end{acknowledgements}

%
%

\bibliographystyle{aa} 
\bibliography{references.bib} 

\begin{appendix} 

\section{Generating the Datasets}
\label{sec:datasets}
For deep learning, it is necessary to have a huge and realistic dataset to obtain relevant results. The codes used to generate our datasets are written in \texttt{Python 3.6}, use \texttt{GalSim} \citep{galsim2015}, \texttt{TensorFlow 1.15.2}, and are publicly available on GitHub (see section \ref{sec:reproducible_research}). The noise is added on the go and is adjusted according to the application. 

\subsection{The Optical Dataset}
\label{subsec:optical_data}
    We developed the CFHT2HST dataset such that the target images are \acrshort{HST}-like and the input images are CFHT-like. The dataset is constituted of the following:
    		
    \begin{itemize}
    	\item \emph{Target Galaxy} - Also called \emph{ground truth galaxy}. This galaxy is obtained by convolving a real \acrfull{HST} galaxy image from the COSMOS catalog \citep{cosmos2012mandelbaum} with the target \acrshort{PSF} to reach a determined target resolution.
    	
    	\item \emph{Target \acrshort{PSF}} - The role of this \acrshort{PSF} is to limit the resolution of the target galaxy to obtain a realistic simulation of an image observed by a telescope. The target \acrshort{PSF} is unique since its variations are negligible with respect to the input \acrshort{PSF}.
    	
    	\item \emph{Input Galaxy} - The image is obtained by convolving the image from the COSMOS catalog, with the input \acrshort{PSF}, and adding a noise with a constant standard deviation for all images. The value of the standard deviation is computed by taking into account the conditions of observation and the properties of the \acrshort{CFHT} telescope. \href{https://github.com/CosmoStat/ShapeDeconv/blob/master/data/CFHT/HST2CFHT.ipynb}{\faFileCodeO}
    	
    	\item \emph{Input \acrshort{PSF}} - This \acrshort{PSF} is used to obtain the input galaxy. It has been generated using a Kolmogorov model \citep{racine1996telescope}.
    \end{itemize}
     
    The characteristics of this dataset are given in Table \ref{table:cfht2hst_dataset}.
    
    \begin{table}[h]
    	\centering
    	\begin{tabular}{|p{4cm}|p{4cm}|} 
    		\hline
    		Pixel scale & $0.187^{\prime \prime}$ \\
    		\hline
    		Image dimensions & $64 \times 64$ \\
    		\hline
    		No. of objects & $51$~$000$ \\
    		\hline
    		Noise standard deviation & $23.59$ \\
    		\hline
    	\end{tabular}
    	\caption{Characteristics of the CFHT2HST dataset.}
    	\label{table:cfht2hst_dataset}
    \end{table}

\subsection{The Radio Dataset}
\label{subsec:radio_data}

    To our knowledge, a public radio dataset of this magnitude and suitable for our working environment does not exist. Hence, we have developed our own dataset composed of over 50,000 images with realistic \acrshort{PSF}s similar to those of the MeerKAT telescope (a precursor to the \acrshort{SKA}) at a central frequency of 3600 MHz, and parametric galaxies with realistic properties taken from the \acrshort{T-RECS} catalog \citep{bonaldi2019}. The key point to note here is that the noise is masked by the \acrshort{PSF} in Fourier space. The noise level is measured in terms of the peak SNR (noted PSNR), and defined as the ratio of the maximum useful signal to the standard deviation of the noise (a commonly used convention in radio astronomy). 
    
    The generated dataset contains pairs of galaxies and PSFs, the characteristics of which are in Table \ref{tab:radio_data}. We use simulated realizations of observational \acrshort{PSF}s from the MeerKAT telescope for typical observation times of 2 hours. The \acrshort{PSF} is completely determined by: i) the observation wavelength, ii) the integration time of the observation, iii) the distribution of the antennas as seen from the source during the observation. To simulate a variety of realistic cases, we generated integrated PSF realizations over 2 hours for random source directions. A longer observation time allows to accumulate more varied samples in the Fourier plane, which decreases artifacts due to incompleteness of the sample mask.

    \begin{table}[h]
        \centering
        \begin{tabular}{|c|c|c|c|}
        \hline
        \multicolumn{4}{|c|}{\textbf{Common Characteristics}} \\
        \hline
            \multicolumn{2}{|c|}{Pixel Scale} & \multicolumn{2}{|c|}{0.58"} \\ 
            \multicolumn{2}{|c|}{Central Frequency} & \multicolumn{2}{|c|}{3600 MHz} \\
            \multicolumn{2}{|c|}{Image Dimensions} & \multicolumn{2}{|c|}{128 X 128} \\
            \multicolumn{2}{|c|}{No. of objects} & \multicolumn{2}{|c|}{51000} \\
        \hline
        \multicolumn{2}{|c|}{\textbf{Galaxy}} & \multicolumn{2}{|c|}{\textbf{PSF}} \\ 
        \hline
            Type & SFG & Type & MeerKAT\\
            Profile & Exponential & No. of Antennas & 64 \\
            Min. size & 6.5 pixels & Observation Time & 2 hours \\
            Max. size & 80 pixels & Time Step & 300 seconds \\
        \hline
        \end{tabular}
        \caption{Characteristics of the simulated MeerKAT3600 dataset.}
        \label{tab:radio_data}
    \end{table}


\section{Implementation}
\label{sec:implementation}
\subsection{Optical Dataset}
\label{subsubsec:implementation_optical}
    
    For the optical experiments, we considered the CFHT2HST dataset presented in section \ref{subsec:optical_data}. In the same spirit as the experiments carried out in \citep{sureau2020}, the goal is to reconstruct high resolution images from low resolution images, with each resolution corresponding to a telescope. In our case, the high resolution images correspond to real images from the \acrshort{HST} telescope and the low resolution images correspond to simulated images at the resolution of the \acrshort{CFHT} obtained by degrading the \acrshort{HST} images. These are considered the ground truths and the \acrshort{CFHT} images are the observed images. To measure the quality of the signal in the image, we use the absolute magnitude. Indeed, the noise level added to the latter is a constant background noise which is calculated from the parameters of the telescope \footnote{For more details on generation of the observed galaxies and calculation of the noise level see: \url{https://github.com/CosmoStat/ShapeDeconv/blob/master/data/CFHT/HST2CFHT.ipynb}}. Therefore, the absolute magnitude is, up to a multiplicative constant, the opposite of the logarithm of the \acrshort{SNR}. We considered 4 classes of magnitudes, each having an average of 20.79, 22.16, 22.83, and 23.30, and each containing an equal number of samples. The dataset contains point-like galaxies, very small to be resolved by the telescope. Since measuring the shape of these galaxies is problematic, we removed from our analysis galaxies where shape measurement on the ground truth image had failed. In each class of magnitudes, then we have about 500 galaxies. The studied methods are \acrshort{SRA} and \acrshort{SCORE} for sparse methods and Tikhonet and ShapeNet, for deep learning. In these last two, the U-net used contains 4 scales and its training took place over 10 epochs composed of 625 steps each with batches having a size of 128 each. For the choice of the shape constraint weight in the ShapeNet method, we performed a linear search and found the value $ \gamma = $ 0.5. In the case of \acrshort{SCORE}, the weight has been set to $ \gamma = 1 $, based on our previous findings in \cite{nammour2021galaxy}. In addition, the convolution kernel used in \acrshort{SCORE} is obtained by performing a division, in Fourier space, of the transform of the input \acrshort{PSF} by the output one. We then perform a \emph{partial deconvolution}.
    
\subsection{Radio Dataset}
\label{subsubsec:implementation_radio}
    
    For the radio experiments, we used the MeerKAT3600 dataset presented in \ref{subsec:radio_data}. Unlike the optical case, ground truth radio images are realistic but are not real. They are simulated images using the \acrshort{T-RECS} catalog \citep{bonaldi2019} and their resolution is not limited by a telescope's dirty beam. However, observations are simulated so that they are similar to those of the MeerKAT telescope. To do so, we used the realistic simulation code that we developed using galaxy2galaxy \footnote{For more details on simulating realistic MeerKAT images with the T-RECS catalog, see: \url{https://github.com/CosmoStat/ShapeDeconv/blob/master/data/T-RECS/Generate\%20Radio\%20ground\%20truth\%20from\%20T-RECS.ipynb}}. The noise level used for these experiments is constant and is chosen so as to obtain a variety of levels of \acrshort{SNR} in order to have a broad assessment of the different methods examined. We use \acrshort{PSNR} to quantify the signal quality. We considered 4 \acrshort{PSNR} classes each with an average of 4.38, 6.92, 9.99, and 16.74, containing an equal number of samples. The methods compared are CLEAN isotropic, \acrshort{SRA}, \acrshort{SCORE}, Tikhonet and ShapeNet. Note that we only consider CLEAN isotropic in our comparisons (and denote it as CLEAN) since it is an improvement to the original CLEAN algorithm and allows a fairer comparison to be made with other methods. The dataset contains observations with a very low PSNR going below 3 which corresponds to the signal detection threshold for CLEAN. In the subset chosen to perform numerical experiments, 72 out of 3072 galaxies were removed because their \acrshort{PSNR} was below the threshold. In the end, we had 750 galaxies per class of \acrshort{PSNR}. For the deep learning, the U-Net used contains 4 scales and was trained for 10 epochs, having 3125 steps per epoch, with a batch size of 32. For the choice of weighting for the shape constraint, we found the value $\gamma = 0,5$ for ShapeNet and $\gamma=2$ for \acrshort{SCORE} using a linear search. The value of the regularization weight for the Tikhonov filter used is $9 \times 10^{-3}$ and was also found using a linear search (link to the notebook: \href{https://github.com/CosmoStat/ShapeDeconv/blob/master/notebooks/Tikhonov_filter/WienerParameterSearch.ipynb}{\faGithub})  Moreover, for these experiments, we also modified the initialization of the sparse methods, \acrshort{SRA} and \acrshort{SCORE}, by replacing the constant image by a Tikhonov filtering applied to the observation. 

\section{Reproducible Research}
\label{sec:reproducible_research}
\begin{enumerate}
  \item The branch of the GitHub repository for:
  \begin{itemize}
    \item Optical dataset generation using galaxy2galaxy: \href{https://github.com/fadinammour/galaxy2galaxy/tree/cfht2hst_prblm}{\faGithub} 
    \item Radio dataset generation using galaxy2galaxy: \href{https://github.com/fadinammour/galaxy2galaxy/tree/radio_data}{\faGithub} 
  \end{itemize}
  \item Scripts for building and training Tikhonet \& ShapeNet: \href{https://github.com/CosmoStat/ShapeDeconv/blob/master/scripts/tikhonet/tikhonet_train_radio_64.py}{\faGithub} 
  \item Evaluating the trained network for different shape constraint parameters: \href{https://github.com/CosmoStat/ShapeDeconv/blob/master/notebooks/UNET_Evaluation/radio/unet_eval_64_multiple_gamma_shape_investigation.ipynb}{\faGithub}
  \item Link to the trained network weights: \href{https://doi.org/10.5281/zenodo.5552714}{\faFileCodeO}
\end{enumerate}

\end{appendix}

\end{document}